# Tunable Flux Qubit manipulated by fast pulses: operating requirements, dissipation and decoherence

F. Chiarello

*Istituto di Fotonica e Nanotecnologie CNR, via del Cineto Romano 42, 00156 Rome, Italy*
*fabio.chiarello@ifn.cnr.it*

**Abstract**

A double SQUID manipulated by fast magnetic flux pulses can be used as a tunable flux qubit. In this paper we study the requirements for the qubit operation, and evaluate dissipation and decoherence due to the manipulation for a typical system. Furthermore, we discuss the possibility to use an integrated Rapid Single Flux Quantum logic for the qubit control.

## 1. Introduction

Quantum computing can overcome the intrinsic limitations of classical computing, and it is a formidable framework for the study and understanding of quantum mechanics. Different qubits (the basic elements of a quantum computer) based on solid state superconducting devices have been realized and tested, individually and in simple coupled configurations [1-14]. Their coherent manipulation, generally performed by NMR-like microwave excitations or by fast pulses, is a critical question: it must be fast, reliable, clean, simple and easily integrable. In this direction one of the more interesting possible strategies consists in the use of an integrated Rapid Single Flux Quantum (RSFQ) logic for the qubit manipulation [15-17].

In this paper we consider a particular qubit based on a Superconducting Quantum Interference Device (SQUID) with a high degree of tunability, the so called double SQUID [18-22], manipulated by fast pulses of magnetic flux. The behaviour of a typical device of this kind is studied in order to define its operating parameters, to fix the requirement for the manipulating pulses, and to evaluate dissipation and decoherence due to the manipulation. Finally it is considered the possibility to use RSFQ logic with appropriate modifications in order to perform the qubit manipulation.

## 2. The tunable flux qubit

A rf SQUID consists of a superconducting loop of inductance $L$, interrupted by a Josephson junction of critical current $I_0$ and capacitance $C$, and biased by an applied magnetic flux $\Phi_x$. For appropriate conditions it can be effectively used as a flux qubit. A more tunable device, the double SQUID [18,19], is obtained by replacing the single junction with a dc SQUID, a smaller superconducting loop of inductance $l$ interrupted by two identical junction of critical current $J$ and capacitance $C_0$ each, biased by an applied magnetic flux $\Phi_c$ (Fig. 1a). For $l \ll \varphi_0/J$ (where $\Phi_0 \cong 2.07 \times 10^{-15} Wb$ is the flux quantum and $\varphi_0 = \Phi_0/2\pi$ is the reduced flux quantum) the dc SQUID behaves approximately like a single junction with total capacitance $C = 2C_0$ and tunable critical current $I_0 = 2J\cos(\pi \Phi_c/\Phi_0)$, and the double SQUID can be used as a tunable rf SQUID. The system dynamics is described by the canonical variable $\varphi$ (the phase difference across the dc SQUID, related to the flux $\Phi$ in the large loop by $\boldsymbol{j} = \Phi/\boldsymbol{j}_0$), and by the relative conjugate variable $p = -i\hbar\partial/\partial\boldsymbol{j}$, with Hamiltonian:

$$H = \frac{p^2}{2M} + E_L\left[\frac{1}{2}(\boldsymbol{j}-\boldsymbol{j}_x)^2 - \boldsymbol{b}\cos(\boldsymbol{j})\right] \quad (1)$$

where $E_L = \boldsymbol{j}_0^2/L$ is the energy scale, $\boldsymbol{j}_x = \Phi_x/\boldsymbol{j}_0$ and $\boldsymbol{j}_c = \Phi_c/\boldsymbol{j}_0$ are the reduced control fluxes, $M = C\boldsymbol{j}_0^2$ is the effective mass, and $\boldsymbol{b} = 2JL/\boldsymbol{j}_0 \cos(\boldsymbol{j}_c/2)$.

For $\boldsymbol{j}_x = \boldsymbol{p}$ (corresponding to $\Phi_x = \Phi_0/2$) the potential is symmetric, with two identical minima separated by a barrier if it is also $1 < \boldsymbol{b} < 4.60$ (Fig. 1b). In this case the energy spectrum is characterized by a degenerate situation, with the first two levels separated by an energy gap $\hbar\Delta = E_1 - E_0$ that is smaller than the separation from upper levels ($E_2 - E_1 \gg \hbar\Delta$). In the absence of possible excitation to these upper levels (due for example to the temperature or to nonadiabatic modifications) a two state approximation can be used by considering the reduced energy base with just the first two energy eigenstates $|0\rangle$ and $|1\rangle$. A second base can be used, consisting of the two flux states centred in the left and right minima respectively, with approximately $|L\rangle = (|0\rangle + |1\rangle)/\sqrt{2}$ and $|R\rangle = (|0\rangle - |1\rangle)/\sqrt{2}$. Also in the asymmetric case, for $\boldsymbol{j}_x$ different but close to $\boldsymbol{p}$, one can again use the two state approximation; now the Hamiltonian (1) in the flux base can be rewritten as follows:

$$H_{flux} = -\frac{\hbar\Delta}{2}\boldsymbol{s}_x - \frac{\hbar\boldsymbol{e}}{2}\boldsymbol{s}_z \quad (2)$$

where $\boldsymbol{s}_x, \boldsymbol{s}_y, \boldsymbol{s}_z$ are the standard Pauli matrices, and $\hbar\boldsymbol{e}$ is the energy separation between the two minima (potential asymmetry) (Fig. 1c). The eigenstates of Hamiltonian (2) can be written as $|\tilde{0}\rangle = c|L\rangle + s|R\rangle$ and $|\tilde{1}\rangle = s|L\rangle - c|R\rangle$, with $c = \cos(\boldsymbol{q}/2)$, $s = \sin(\boldsymbol{q}/2)$, and $\boldsymbol{q} = \arctan(\Delta/\boldsymbol{e})$, while the energy gap between these states is $\hbar\Omega = \hbar\sqrt{\Delta^2 + \boldsymbol{e}^2}$. We observe that the states $|\tilde{0}\rangle$ and $|\tilde{1}\rangle$ are equivalent to $|0\rangle$ and $|1\rangle$ just in the symmetric case, when $\boldsymbol{e} = 0$ and therefore $\boldsymbol{q} = \boldsymbol{p}/2$.

The considered system can be used as a qubit by mapping the computational qubit states "0" and "1" in, for example, the two distinct flux states $|L\rangle$ and $|R\rangle$. The possibility to tune the parameter $\Delta$, generally fixed in other kind of superconducting qubits, allows a complete control of the qubit and justifies the name "tunable qubit" used for this system. NMR-like manipulation with microwave pulses can be performed like for other superconducting qubits, but now it is also possible a complete manipulation just with fast flux pulses. For example, the state preparation can be done by strongly unbalancing the potential with the control $\Phi_x$ in order to have just one minimum, then waiting a time sufficient for the relaxation to this minimum, and finally returning to the symmetric situation still maintaining the barrier high, in a state that is "frozen". Coherent rotation between the two states can be achieved by reducing the barrier in order to have fast free oscillations, then waiting the time necessary for the desired rotation, and finally raising again the barrier to return in the frozen state. Other kind of manipulations can be performed with similar sequences of variations, allowing the full control of the qubit [20-22].

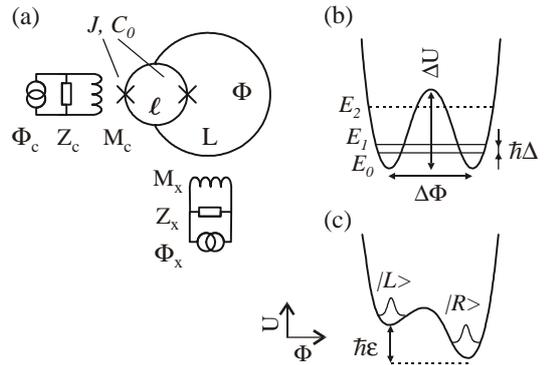

Fig. 1 (a) Scheme of the double SQUID with the two control coils. (b) Potential of the double SQUID in the symmetric case, and relative energy levels. (c) Potential in the asymmetric case.

## 3. Operation requirements

In this paragraph we study, both numerically and with analytical approximations, the behaviour of the tunable flux qubit in order to derive the requirement on the control pulses for both $\Phi_x$ and $\Phi_c$ for typical qubit parameters.

We introduce some important parameters in the symmetric case: the distance $\Delta j$ between the minima, the barrier height $\Delta U$, and the small oscillation frequency in the minima $w_b = \sqrt{(dU^2/d^2 j)_{\min}/M}$ (Fig. 1b). The system dynamics is interesting just for $b$ greater and close to 1, so that these parameters can be expanded in series for $0 < b - 1 \ll 1$ in order to derive approximated analytical expressions. A correction, obtained by multiplying the results of the expansions for appropriate powers of $b$ empirically determined, can be used in order to extend the validity in the range $1 < b < 2$ with a $^0/_{00}$ accuracy:

$$\Delta j \cong \sqrt{24(b-1)}\, b^{-0.36}$$
$$\Delta U \cong \frac{3}{2} E_L (b-1)^2 b^{-0.82} \quad (3)$$
$$w_b \cong w_L \sqrt{2(b-1)}\, b^{-0.145}$$

where $w_L = 1/\sqrt{LC}$. The parameter $\Delta$ can be evaluated by using an approximated expression in the limit $\Delta U \gg \hbar w_b$ [23]:

$$\Delta \cong A w_b \sqrt{B \frac{\Delta U}{\hbar w_b}} \exp\left(-B \frac{\Delta U}{2\hbar w_b}\right) \quad (4)$$

with $A \cong 1$ and $B \cong 10.2$. In a more general case, considering also a slightly asymmetric potential with energy unbalancing $\hbar e = E_L j_x \Delta j$ (Fig. 1c), we have $E_1 - E_0 = \hbar \Omega = \hbar \sqrt{\Delta^2 + e^2}$, and the spacing $E_2 - E_1$ can be roughly estimated with $\hbar(w_b - \Omega)$. These analytical approximated results can be compared with numerical evaluations, obtained by solving the time independent Schrodinger equation with Hamiltonian (1) using standard numerical techniques. Let us consider a realistic case by choosing a set of typical parameters for the double SQUID: large loop inductance $L = 85\,pH$, small loop inductance $l < 5\,pH$, single junction critical current $J = 5\,mA$ and capacitance $C_0 = 0.25\,pF$. In Fig. 2 we plot the level spacing $\Delta/2p = (E_1 - E_0)/h$ in the symmetric case as a function of the flux control $\Phi_c$, obtained both analytically (lower straight line) and numerically (lower dashed line), and the spacing to upper levels $(E_2 - E_1)/h$, again obtained both analytically (upper straight line) and numerically (upper dashed line). These curves can be used to fix the fundamental requirements on the control pulse $\Phi_c$. Manipulations are performed by considering the switching between two distinct working points, a frozen state (F) where the barrier is very high, and an evolving state (E) where the barrier is low and the free evolution occurs. We choose (F) and (E) in order to have $\Delta_F/2p \approx 100\,kHz$ and $\Delta_E/2p = 500\,MHz$ respectively, obtaining $\Phi_{cF} = 359.3\,m\Phi_0$ and $\Phi_{cE} = 367.6\,m\Phi_0$, corresponding to a pulse amplitude $\Delta\Phi_c = 8.2\,m\Phi_0$. The rise/fall time $t$ of this pulse must be chosen with some attention. In fact the variation rate must be fast with respect to the free evolution frequency $\Delta/2p = 500\,MHz$, but at the same time it must not excite upper non computational states, and so it must also be smaller than $(E_2 - E_1)/h \approx 4\,GHz$ (in the point E, the worst case), so that it must be $100\,ps < t < 800\,ps$.

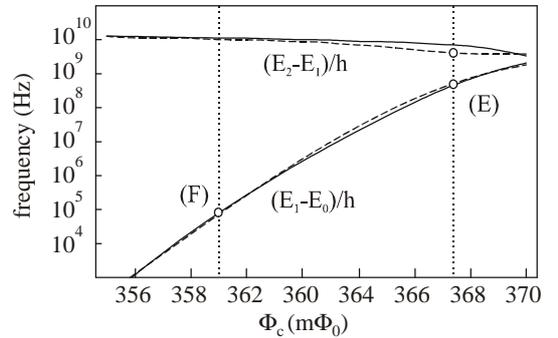

Fig. 2 Energy levels spacing $(E_1 - E_0)/h$ (lower curves) and $(E_2 - E_1)/h$ (upper curves), obtained both analytically (straight lines) and numerically (dashed lines) respectively.

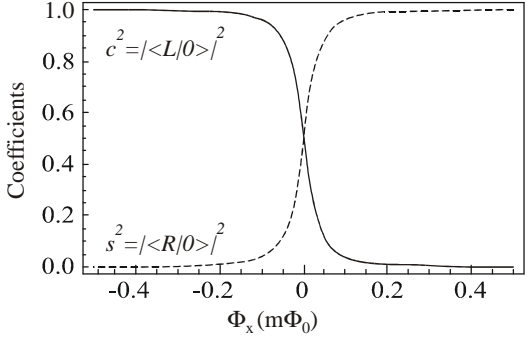

Fig. 3 Coefficients $c^2$ and $s^2$, obtained by the squared projections of the left (straight line) and right (dashed line) states on the fundamental state.

In Fig. 3 there are plotted the coefficients $c^2 = \left|\langle L|\tilde{0}\rangle\right|^2 = (1+\boldsymbol{e}/\Omega)/2$ (straight line) and $s^2 = \left|\langle R|\tilde{0}\rangle\right|^2 = (1-\boldsymbol{e}/\Omega)/2$ (dashed line) as functions of $\Phi_x$ for $\Phi_c = \Phi_{cE}$. From these curves it results that a complete preparation/manipulation of the flux state requires a variation from the symmetric point $\Phi_x = \Phi_0/2$ larger than $\Delta\Phi_x \approx 0.2\,m\Phi_0$. The rise/fall times of the $\Phi_x$ pulse must undergo the same requirements as the $\Phi_c$ pulse.

## 4. Dissipation and decoherence

In this paper we are just interested in studying dissipation and decoherence due to the manipulating flux pulses on $\Phi_x$ and $\Phi_c$. These controls are applied by using superconducting coils of inductance $L_k$ (the index $k = x, c$ identifies which of the two controls we are considering) coupled with mutual inductance $M_k$ to the considered qubit loop. Each coil is biased by a circuit that we model with an ideal current source $I_k$ in parallel to a frequency dependent complex impedance $Z_k(\boldsymbol{w})$ (Fig. 1a), and to the related generator of current noise $\boldsymbol{d}I_k$ with spectral density $S_{dI_k} = 4k_b\tilde{T}_k/\mathrm{Re}(Z_k)$, where $\tilde{T}_k(\boldsymbol{w}) = \hbar\boldsymbol{w}/2k_b \coth(\hbar\boldsymbol{w}/2k_bT)$ is the effective temperature. The current noise $\boldsymbol{d}I_k$ causes a corresponding flux noise $\boldsymbol{d}\Phi_k = M_k \boldsymbol{d}I_k$, and a consequent fluctuation of the parameters $\Delta$ and $\boldsymbol{e}$.

For small noise contributions, in linear approximation, we can assume $\boldsymbol{d}\Delta = (d\Delta/d\Phi_c)\boldsymbol{d}\Phi_c$ and $\boldsymbol{d}\boldsymbol{e} = (d\boldsymbol{e}/d\Phi_x)\boldsymbol{d}\Phi_x$ respectively, with spectral densities that can be rearranged in the following expressions:

$$S_{d\Delta}(\boldsymbol{w}) = \frac{4k_b\tilde{T}_c}{\hbar}\frac{R_0}{R_c}\left(\frac{M_c}{L}\right)^2\left[\frac{\partial(\hbar\Delta/E_L)}{\partial(\Phi_c/\boldsymbol{j}_0)}\right]^2$$

$$S_{d\boldsymbol{e}}(\boldsymbol{w}) = \frac{4k_b\tilde{T}_x}{\hbar}\frac{R_0}{R_x}\left(\frac{M_x}{L}\right)^2\left[\frac{\partial(\hbar\boldsymbol{e}/E_L)}{\partial(\Phi_x/\boldsymbol{j}_0)}\right]^2 \quad (5)$$

with $R_0 = \boldsymbol{j}_0^{\,2}/\hbar \cong 1026\,\Omega$ and $R_k = \mathrm{Re}[Z_k(\boldsymbol{w})]$. The noise contributions can be added in Hamiltonian (2) giving:

$$H_{flux} = \left(-\frac{\hbar\Delta}{2}\boldsymbol{s}_x - \frac{\hbar\boldsymbol{e}}{2}\boldsymbol{s}_z\right) - \hbar\,\boldsymbol{d}\Delta\,\boldsymbol{s}_x - \hbar\,\boldsymbol{d}\boldsymbol{e}\,\boldsymbol{s}_z \quad (6)$$

In the energy base of the noiseless Hamiltonian these contributions can be reorganized in a longitudinal and in a transverse part:

$$H_{energy} = -\frac{\hbar\Omega}{2}\boldsymbol{s}_z - \hbar\left(\boldsymbol{d}\boldsymbol{e}\,\frac{\boldsymbol{e}}{\Omega} + \boldsymbol{d}\Delta\,\frac{\Delta}{\Omega}\right)\boldsymbol{s}_z - \\ -\hbar\left(\boldsymbol{d}\boldsymbol{e}\,\frac{\Delta}{\Omega} - \boldsymbol{d}\Delta\,\frac{\boldsymbol{e}}{\Omega}\right)\boldsymbol{s}_x \quad (7)$$

The standard two state system theory for a small perturbing noise in the form of equation (7) gives simple expressions for the relaxation rate $\Gamma_1$, the pure dephasing rate $\Gamma_j$ and the dephasing rate $\Gamma_2$ [24]:

$$\Gamma_1 = \left(\frac{\Delta}{\Omega}\right)^2 S_{d\boldsymbol{e}}(\Omega) + \left(\frac{\boldsymbol{e}}{\Omega}\right)^2 S_{d\Delta}(\Omega)$$

$$\Gamma_j = \left(\frac{\boldsymbol{e}}{\Omega}\right)^2 S_{d\boldsymbol{e}}(0) + \left(\frac{\Delta}{\Omega}\right)^2 S_{d\Delta}(0) \quad (8)$$

$$\Gamma_2 = \frac{1}{2}\Gamma_1 + \Gamma_j$$

We notice from equations (5) and (8) that relaxation and decoherence depends quadratically on the coupling strengths $(M_k/L_k)$ and only linearly on the effective temperatures and on the dissipating contributions $R_k/R_0$. This gives an important guideline for the design of the qubit control, indicating the convenience of a strong decoupling from the control circuit.

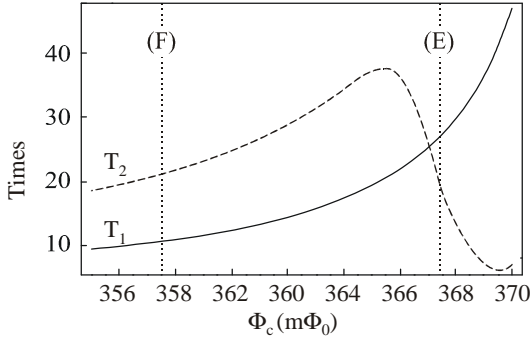

Fig. 4 Times $T_1$ and $T_2$ as a function of the control flux $\Phi_c$.

Let us consider typical control current pulses with amplitude of the order of $10\,mA$. In order to generate the required flux pulses $\Delta\Phi_x = 0.2\,m\Phi_0$ and $\Delta\Phi_c = 10\,m\Phi_0$ we need couplings $M_x = 41\,fH$ and $M_c = 1.7\,pH$.

For a first rough evaluation we assume $R_k = 50\,\Omega$ for any frequency, and $T = 10\,mK$. With these assumptions and by using equation (8) we can plot $T_1 = \Gamma_1^{-1}$ (straight line) and $T_2 = \Gamma_2^{-1}$ (dashed line) in the symmetric case as a function of $\Phi_c$ (Fig. 4). The obtained values $T_1$ and $T_2$, well above $10\,ms$ during all possible manipulations, must be compared with the typical time require for one operation (in our case below $2\,ns$), so that it results from this rough evaluation that the considered system and manipulation procedure is in principle adequate for quantum computing applications. This result can be simply scaled for different assumptions by using equations (5), and it is also possible to introduce more complex, not flat noise spectral densities.

## 5. RSFQ manipulation

RSFQ logic is an architecture based on resistively shunted Josephson elements, allowing ultra-fast digital operations [25]. It can be simply and effectively used in order to produce the flux pulses necessary for the qubit manipulation described in this paper. For example, in a RSFQ flip-flop there is a superconducting loop in which it is stored a flux that can switch in a controllable manner between two distinct states differing just for one flux quantum. Two of these flip-flops can be coupled to the two qubit loops by means of two different superconducting transformers that ensure the required couplings for the manipulation [15-17]. Two successive switchings in one of the flip-flop generates the flux pulse that manipulates the qubit as previously described. All the required devices, qubits and RSFQ controls, can be realized in the same chip using the same technology, with great advantage in the simplicity and integrability of the system, but one can also use coupled-chip design if necessary.

The operation requirements defined in this paper are in part suitable for the integration with RSFQ control logic, but there are some problems. First of all, since the RSFQ logic requires resistive shunts, the typical impedance seen by the qubit is very small, of the order of few Ohms or less, with consequent reduction of times $T_1$ and $T_2$. Second, typical RSFQ pulse rise/fall times are extremely short, of the order of tens of picoseconds. If directly applied to the qubit, these signals induce transitions to non computational states. Finally, standard RSFQ circuitry it is designed to work at Helium temperature (4.2K), but for quantum computing applications lower temperature (10mK) is needed.

These problems can be solved by developing an unconventional RSFQ logic [26], together with an appropriate filtering of the transmitted signal and an optimization of the qubit parameters. Different efforts are starting in this direction, in particular in the frame of the UE project "RSFQubit", and first prototypes of chips with an RSFQ flip-flop coupled to a tunable flux qubit are under fabrication. The first results will give important indication for the future developments.

## 6. Conclusions

A double SQUID with a flux pulse control scheme can be effectively used as a tunable flux qubit. In this paper we study the requirements of the control fluxes for a device with typical parameters, both with approximated analytical expressions and with numerical simulation, obtaining pulses amplitudes of the two controls $\Delta\Phi_x \approx 0.2\,m\Phi_0$ and $\Delta\Phi_c \approx 8\,m\Phi_0$, with rise/fall time $t$ that must be in the limit $100\,ps < t < 800\,ps$. Relaxation and decoherence times (just due to the manipulating pulses) are expected to be well above $10\,\mathbf{\mathit{ms}}$ if one suppose an effective dissipation of $50\Omega$ at $10\,mK$, that is suitable for quantum computing applications.

RSFQ logic is an interesting candidate for the control of this qubit, provided the development of an unconventional RSFQ design that would accomplish the qubit requirements. Possible future work concerns the optimization of the parameters for the considered system, the development and test of an RSFQ qubit control, and the realization of a final experiment for the observation of coherent manipulation with pulses in a tunable flux qubit.

**Acknowledgements**

We thank M. G. Castellano, G. Torrioli, C. Cosmelli, P. Carelli, A. Zorin and M. Khabipov for the useful discussions. Thanks are also due to M. Castagna. This work was supported in part by "RSFQubit" FP6 project of European Union, an by the INFN "SQC" project.